\documentclass[english,a4paper,12pt]{article}
\usepackage{varioref}
\usepackage{latexsym}
\usepackage{amsmath}
\usepackage{tabularx}

\numberwithin{equation}{section}
\input{tcilatex}

\begin{document}

\title{Some first thoughts on the stability of the asynchronous systems}
\author{SERBAN\ VLAD \\
Department of computers, \\
Oradea City Hall, Oradea, Romania\\
web: www.geocities.com/serban\_e\_vlad}
\date{}
\maketitle

\begin{center}
\begin{tabular}{p{12cm}}
\hline
\textbf{Abstract}. {\footnotesize The (non-initialized, non-deterministic)
asynchronous systems (in the input-output sense) are multi-valued functions
from m-dimensional signals to sets of n-dimensional signals, the concept
being inspired by the modeling of the asynchronous circuits. Our purpose is
to state the problem of the their stability.} 
\\ \hline
\end{tabular}
\end{center}

\section{Introduction}

$\mathbf{B}=\{0,1\}$ is the binary Boole algebra. The function $x:\mathbf{R}%
\rightarrow \mathbf{B}^{n}$ has a limit when $t\rightarrow \infty $ if%
\begin{equation}
\exists t_{f},\forall t\geq t_{f},x(t)=x(t_{f})
\end{equation}%
The usual notation is $x(t_{f})=\underset{t\rightarrow \infty }{\lim }x(t)$. 
$x$ is called (n-dimensional) signal if it is of the form%
\begin{equation}
x(t)=x(t_{0}-0)\cdot \varphi _{(-\infty ,t_{0})}(t)\oplus x(t_{0})\cdot
\varphi _{\lbrack t_{0},t_{1})}(t)\oplus x(t_{1})\cdot \varphi _{\lbrack
t_{1},t_{2})}(t)\oplus ...
\end{equation}%
where $t\in \mathbf{R}$. In (1.2) $\varphi _{(\quad )}:\mathbf{R}\rightarrow 
\mathbf{B}$ is the characteristic function and $t_{0}<t_{1}<t_{2}<...$ is
some unbounded sequence. We note%
\begin{equation*}
S^{(n)}=\{x|x:\mathbf{R}\rightarrow \mathbf{B}^{n},x\text{ }is\text{ }%
signal\}
\end{equation*}%
\begin{equation*}
P^{\ast }(S^{(n)})=\{X|X\subset S^{(n)},X\neq \emptyset \}
\end{equation*}%
\begin{equation*}
S_{c}^{(n)}=\{x|x\in S^{(n)},\exists \underset{t\rightarrow \infty }{\lim }%
x(t)\}
\end{equation*}%
For the Boolean function $F:\mathbf{B}^{m}\rightarrow \mathbf{B}^{n}$ we
note also%
\begin{equation*}
S_{F,c}^{(m)}=\{u|u\in S^{(m)},\exists \underset{t\rightarrow \infty }{\lim }%
F(u(t))\}
\end{equation*}%
Any signal $x$ has an initial time instant $t_{0}$, from the definition
(1.2). It is not unique and it is precised by the condition $\forall
t<t_{0},x(t)=x(t_{0}-0)$, where (the unique) $x(t_{0}-0)$ is the initial
value of $x$. In particular the constant function $x$ satisfies the property
that any $t_{0}$ is an initial time instant and $x$ coincides with its
initial value. There exist signals without final time instant $t_{f}$ and
respectively without final value $\underset{t\rightarrow \infty }{\lim }x(t)$%
. If $t_{f}$ exists, it is not unique and any $t_{f}^{\prime }>t_{f}$ is a
final time instant too. In particular, the constant function $x$ satisfies
the property that any $t_{f}$ is a final time instant and $x$ coincides with
its final value.

When $x$ is the state of a system, the problem of the existence of $t_{f}$,
thus of the limit $\underset{t\rightarrow \infty }{\lim }x(t)$ is the
stability problem of that system.

\section{Asynchronous systems}

\textbf{Definition} We call (non-initialized, non-deterministic)
asynchronous system (in the input-output sense) a function $f:U\rightarrow
P^{\ast }(S^{(n)}),$ where $U\in P^{\ast }(S^{(m)})$. The elements $u\in U$,
respectively $x\in f(u)$ are called (admissible) inputs, respectively
(possible) states, or outputs.

\textbf{Remark} The concept of asynchronous system has its origin in the
modeling of the asynchronous circuits, where the multivalued association
between the cause $u$ and the effects $x\in f(u)$ is motivated by the
changes in power supply, temperature, by the technologycal dispersion, by
the errors of the measurement instruments etc.

\textbf{Definition} The system $g:V\rightarrow P^{\ast }(S^{(n)}),$ $V\in
P^{\ast }(S^{(m)})$ is called a subsystem of $f$ if%
\begin{equation*}
V\subset U\quad and\quad \forall u\in V,g(u)\subset f(u)
\end{equation*}

\textbf{Defnition} The system $f^{\ast }:U^{\ast }\rightarrow P^{\ast
}(S^{(n)}),$ $U^{\ast }\in P^{\ast }(S^{(m)})$ is called the dual system of $%
f$ if $U^{\ast }=\{\overline{u}|u\in U\}$ and $\forall u\in U,f^{\ast }(%
\overline{u})=\{\overline{x}|x\in f(u)\}$. We have noted with $\overline{u},%
\overline{x}$ the coordinatewise complements of these signals, for example $%
\overline{u}(t)=(\overline{u_{1}(t)},...,\overline{u_{m}(t)})$.

\textbf{Definition} We suppose that $U\cap V\neq \emptyset $ and that $%
\forall u\in U\cap V,f(u)\cap g(u)\neq \emptyset .$ The system $f\cap
g:U\cap V\rightarrow P^{\ast }(S^{(n)})$ is defined by%
\begin{equation*}
\forall u\in U\cap V,(f\cap g)(u)=f(u)\cap g(u)
\end{equation*}

\textbf{Definition} The system $f\cup g:U\cup V\rightarrow P^{\ast }(S^{(n)})
$ is defined in the next manner%
\begin{equation*}
\forall u\in U\cup V,(f\cup g)(u)=\left\{ 
\begin{array}{c}
f(u),if\quad u\in U-V \\ 
g(u),if\quad u\in V-U \\ 
f(u)\cup g(u),if\quad u\in U\cap V%
\end{array}%
\right. 
\end{equation*}

\textbf{Definition} Let the system $f^{\prime }:U^{\prime }\rightarrow
P^{\ast }(S^{(n^{\prime })}),U^{\prime }\in P^{\ast }(S^{(m)})$. If $U\cap
U^{\prime }\neq \emptyset $, the parallel connection of $f$ and $f^{\prime }$
is the system $(f,f^{\prime }):U\cap U^{\prime }\rightarrow P^{\ast
}(S^{(n+n^{\prime })})$ defined by%
\begin{equation*}
\forall u\in U\cap U^{\prime },(f,f^{\prime })(u)=\{z|z\in S^{(n+n^{\prime
})},
\end{equation*}%
\begin{equation*}
\forall i\in \{1,...,n+n^{\prime }\},z_{i}=\left\{ 
\begin{array}{c}
x_{i},\quad if\quad i\in \{1,...,n\},x\in f(u)\quad \quad \quad \quad \quad
\\ 
y_{i-n},if\quad i\in \{n+1,...,n+n^{\prime }\},y\in f^{\prime }(u)%
\end{array}%
\right. \}
\end{equation*}

\textbf{Definition} The system $h:X\rightarrow P^{\ast }(S^{(p)}),X\in
P^{\ast }(S^{(n)})$ is given, so that $\forall u\in U,f(u)\cap X\neq
\emptyset $. The serial connection of $h$ and $f$ is the system $h\circ
f:U\rightarrow P^{\ast }(S^{(p)})$ that is defined by%
\begin{equation*}
\forall u\in U,(h\circ f)(u)=\{y|\exists x\in f(u)\cap X,y\in h(x)\}
\end{equation*}

\textbf{Definition }The system $f$ is called non-anticipatory, or causal if%
\begin{equation*}
\forall t_{1}\in \mathbf{R},\forall u\in U,\forall v\in U,u_{|(-\infty
,t_{1})}=v_{|(-\infty ,t_{1})}\Longrightarrow
\end{equation*}%
\begin{equation*}
\Longrightarrow \{x_{|(-\infty ,t_{1})}|x\in f(u)\}=\{y_{|(-\infty
,t_{1})}|y\in f(v)\}
\end{equation*}

\textbf{Definition} The system $f$ is initialized if%
\begin{equation*}
\exists w^{0}\in \mathbf{B}^{n},\forall u\in U,\forall x\in f(u),\exists
t_{0}\in \mathbf{R},\forall t<t_{0},x(t)=w^{0}
\end{equation*}%
If so, the unique vector $w^{0}$ satisfying the previous property is called
the initial state of $f$.

\section{Steady values of the states}

\textbf{Definition} Let the system $f:U\rightarrow P^{\ast }(S^{(n)}),$ $%
U\subset S^{(m)}$. If 
\begin{equation*}
\exists u\in U,\exists x\in f(u),\exists w\in \mathbf{B}^{n},\exists
t_{f}\in \mathbf{R},\forall t\geq t_{f},x(t)=w
\end{equation*}%
then the binary vector $w$ is called the steady value, or the final value,
or the limit when $t\rightarrow \infty $ of the state $x\in f(u)$. In the
special case when 
\begin{equation*}
\exists u\in U,\exists x\in f(u),\exists w\in \mathbf{B}^{n},\forall t\in 
\mathbf{R},x(t)=w
\end{equation*}%
is true, $w$ is called a point of equilibrium of $f$.

\textbf{Remark} For any $u$ and any $x\in f(u)$, if $w=\underset{%
t\rightarrow \infty }{\lim }x(t)$ exists, then it is unique.

\textbf{Notation} For $u\in U$, we note 
\begin{equation*}
\Sigma _{f}(u)=\{w|\exists x\in f(u),w=\underset{t\rightarrow \infty }{\lim }%
x(t)\}
\end{equation*}

\section{Initial time and final time}

\textbf{Definition} We say that the system $f$ has an initial time (instant) 
$t_{0}$ which is

\begin{enumerate}
\item[a)] unbounded if%
\begin{equation*}
\forall u\in U,\forall x\in f(u),\exists t_{0}\in \mathbf{R},\forall
t<t_{0},x(t)=x(t_{0}-0)
\end{equation*}

\item[b)] bounded if%
\begin{equation*}
\forall u\in U,\exists t_{0}\in \mathbf{R},\forall x\in f(u),\forall
t<t_{0},x(t)=x(t_{0}-0)
\end{equation*}

\item[c)] fix (or universal) if%
\begin{equation*}
\exists t_{0}\in \mathbf{R},\forall u\in U,\forall x\in f(u),\forall
t<t_{0},x(t)=x(t_{0}-0)
\end{equation*}
\end{enumerate}

We say that the system $f$ has a final time (instant) $t_{f}$ which is

\begin{enumerate}
\item[a')] unbounded if%
\begin{equation*}
\forall u\in U,\forall x\in f(u)\cap S_{c}^{(n)},\exists t_{f}\in \mathbf{R}%
,\forall t\geq t_{f},x(t)=x(t_{f})
\end{equation*}

\item[b')] bounded if%
\begin{equation*}
\forall u\in U,\exists t_{f}\in \mathbf{R},\forall x\in f(u)\cap
S_{c}^{(n)},\forall t\geq t_{f},x(t)=x(t_{f})
\end{equation*}

\item[c')] fix (or universal) if%
\begin{equation*}
\exists t_{f}\in \mathbf{R},\forall u\in U,\forall x\in f(u)\cap
S_{c}^{(n)},\forall t\geq t_{f},x(t)=x(t_{f})
\end{equation*}
\end{enumerate}

\textbf{Remarks} There are $3\times 3=9$ possibilities of combining the
initial time and the final time for a system.

The next implications are true:%
\begin{equation*}
t_{0}\quad fix\Longrightarrow t_{0}\quad bounded\Longrightarrow t_{0}\quad
unbounded
\end{equation*}%
and the next implications are true also:%
\begin{equation*}
t_{f}\quad fix\Longrightarrow t_{f}\quad bounded\Longrightarrow t_{f}\quad
unbounded
\end{equation*}

\section{Absolute stability}

\textbf{Definition} a) A system $f$ that satisfies%
\begin{equation*}
\forall u\in U,\forall x\in f(u),\exists w\in \mathbf{B}^{n},\exists
t_{f}\in \mathbf{R},\forall t\geq t_{f},x(t)=w
\end{equation*}%
where $w$ and $t_{f}$ depend on $x$ only (thus $\exists w,\exists t_{f}$
commute) is called absolutely stable.

b) If%
\begin{equation*}
\forall u\in U,\exists w\in \mathbf{B}^{n},\forall x\in f(u),\exists
t_{f}\in \mathbf{R},\forall t\geq t_{f},x(t)=w
\end{equation*}%
then $f$ is called absolutely race-free stable, or absolutely
delay-insensitive.

c) We say that $f$ is absolutely constantly stable if it satisfies%
\begin{equation*}
\exists w\in \mathbf{B}^{n},\forall u\in U,\forall x\in f(u),\exists
t_{f}\in \mathbf{R},\forall t\geq t_{f},x(t)=w
\end{equation*}

\textbf{Remarks} The next implications are true:%
\begin{equation*}
f\quad abs\quad const\quad stable\Longrightarrow f\quad abs\quad
race-free\quad stable\Longrightarrow f\quad abs\quad stable
\end{equation*}%
On the other hand if $f$ is absolutely stable, then it defines the system $%
\lim f:U\rightarrow P^{\ast }(S^{(n)})$ by $\forall u\in U,\lim f(u)=\Sigma
_{f}(u)$ and we have identified the binary vector with the constant vector
function. In case of absolute race-free stability, this system is
deterministic, i.e. $\forall u\in U,$ the set $\lim f(u)$ has exactly one
element. If the absolute constant stability of $f$ is true also, then $\lim
f $ is the constant univalued function.

Sometimes it will be useful to write the absolute stability condition under
the form%
\begin{equation*}
\forall u\in U,\forall x\in f(u),\exists w\in \mathbf{B}^{n},\exists
t_{f}\in \mathbf{R},\forall t\geq t_{f},x(t-0)=w
\end{equation*}%
and similarly for the other two cases, showing the fact that $x$ has reached
its final value $w$ sometime before $t_{f}$.

\textbf{Theorem} We suppose that $f:U\rightarrow P^{\ast }(S^{(n)}),$ $%
U\subset S^{(m)}$ is an absolutely stable (an absolutely race-free stable,
an absolutely constantly stable) system and let the systems $g:V\rightarrow
P^{\ast }(S^{(n)}),$ $V\subset S^{(m)}$, $f^{\prime }:U^{\prime }\rightarrow
P^{\ast }(S^{(n^{\prime })}),$ $U^{\prime }\subset S^{(m)}.$ The next
statements are true:

a) If $g\subset f$, then $g$ is absolutely stable (absolutely race-free
stable, absolutely constantly stable)

b) $f^{\ast }$ is absolutely stable (absolutely race-free stable, absolutely
constantly stable)

c) If $U\cap V\neq \emptyset $ and $\forall u\in U\cap V,f(u)\cap g(u)\neq
\emptyset $, then $f\cap g$ is absolutely stable (absolutely race-free
stable, absolutely constantly stable)

d) If $g$ is absolutely stable (absolutely race-free stable, absolutely
constantly stable), then $f\cup g$ is absolutely stable (absolutely
race-free stable, absolutely constantly stable)

e) If $f^{\prime }$ is absolutely stable (absolutely race-free stable,
absolutely constantly stable) and if $U\cap U^{\prime }\neq \emptyset $,
then $(f,f^{\prime })$ is absolutely stable (absolutely race-free stable,
absolutely constantly stable)

\textbf{Theorem} Let the systems $f$ and $h:X\rightarrow P^{\ast }(S^{(p)}),$
$X\subset S^{(n)}$. We suppose that $\forall u\in U,f(u)\cap X\neq \emptyset 
$; then if $h$ is absolutely stable (absolutely constantly stable), we have
that $h\circ f$ is absolutely stable (absolutely constantly stable).

\textbf{Remark} The statement of the previous theorem is false in the case
of absolute race-free stability, in general.

\textbf{Theorem} The next properties are equivalent for the system $f$:

a) absolute stability with unbounded final time:%
\begin{equation*}
\left\{ 
\begin{array}{c}
\forall u\in U,\forall x\in f(u),\exists w\in \mathbf{B}^{n},\exists
t_{f}\in \mathbf{R},\forall t\geq t_{f},x(t)=w \\ 
\forall u\in U,\forall x\in f(u)\cap S_{c}^{(n)},\exists t_{f}\in \mathbf{R}%
,\forall t\geq t_{f},x(t)=x(t_{f})%
\end{array}%
\right. \Longleftrightarrow
\end{equation*}%
\begin{equation*}
\Longleftrightarrow \forall u\in U,\forall x\in f(u),\exists w\in \mathbf{B}%
^{n},\exists t_{f}\in \mathbf{R},\forall t\geq t_{f},x(t)=w
\end{equation*}%
where $w$ and $t_{f}$ depend on $x$ only (thus $\exists w,\exists t_{f}$
commute)

b) absolute stability with bounded final time:%
\begin{equation*}
\left\{ 
\begin{array}{c}
\forall u\in U,\forall x\in f(u),\exists w\in \mathbf{B}^{n},\exists
t_{f}\in \mathbf{R},\forall t\geq t_{f},x(t)=w \\ 
\forall u\in U,\exists t_{f}\in \mathbf{R},\forall x\in f(u)\cap
S_{c}^{(n)},\forall t\geq t_{f},x(t)=x(t_{f})%
\end{array}%
\right. \Longleftrightarrow
\end{equation*}%
\begin{equation*}
\Longleftrightarrow \forall u\in U,\exists t_{f}\in \mathbf{R},\forall x\in
f(u),\exists w\in \mathbf{B}^{n},\forall t\geq t_{f},x(t)=w
\end{equation*}

c) absolute stability with fix final time:%
\begin{equation*}
\left\{ 
\begin{array}{c}
\forall u\in U,\forall x\in f(u),\exists w\in \mathbf{B}^{n},\exists
t_{f}\in \mathbf{R},\forall t\geq t_{f},x(t)=w \\ 
\exists t_{f}\in \mathbf{R},\forall u\in U,\forall x\in f(u)\cap
S_{c}^{(n)},\forall t\geq t_{f},x(t)=x(t_{f})%
\end{array}%
\right. \Longleftrightarrow
\end{equation*}%
\begin{equation*}
\Longleftrightarrow \exists t_{f}\in \mathbf{R},\forall u\in U,\forall x\in
f(u),\exists w\in \mathbf{B}^{n},\forall t\geq t_{f},x(t)=w
\end{equation*}

d) absolute race-free stability with unbounded final time:%
\begin{equation*}
\left\{ 
\begin{array}{c}
\forall u\in U,\exists w\in \mathbf{B}^{n},\forall x\in f(u),\exists
t_{f}\in \mathbf{R},\forall t\geq t_{f},x(t)=w \\ 
\forall u\in U,\forall x\in f(u)\cap S_{c}^{(n)},\exists t_{f}\in \mathbf{R}%
,\forall t\geq t_{f},x(t)=x(t_{f})%
\end{array}%
\right. \Longleftrightarrow
\end{equation*}%
\begin{equation*}
\Longleftrightarrow \forall u\in U,\exists w\in \mathbf{B}^{n},\forall x\in
f(u),\exists t_{f}\in \mathbf{R},\forall t\geq t_{f},x(t)=w
\end{equation*}

e) absolute race-free stability with bounded final time:%
\begin{equation*}
\left\{ 
\begin{array}{c}
\forall u\in U,\exists w\in \mathbf{B}^{n},\forall x\in f(u),\exists
t_{f}\in \mathbf{R},\forall t\geq t_{f},x(t)=w \\ 
\forall u\in U,\exists t_{f}\in \mathbf{R},\forall x\in f(u)\cap
S_{c}^{(n)},\forall t\geq t_{f},x(t)=x(t_{f})%
\end{array}%
\right. \Longleftrightarrow
\end{equation*}%
\begin{equation*}
\Longleftrightarrow \forall u\in U,\exists w\in \mathbf{B}^{n},\exists
t_{f}\in \mathbf{R},\forall x\in f(u),\forall t\geq t_{f},x(t)=w
\end{equation*}%
where $w$ and $t_{f}$ depend on $u$ only (thus $\exists w,\exists t_{f}$
commute)

f) absolute race-free stability with fix final time:%
\begin{equation*}
\left\{ 
\begin{array}{c}
\forall u\in U,\exists w\in \mathbf{B}^{n},\forall x\in f(u),\exists
t_{f}\in \mathbf{R},\forall t\geq t_{f},x(t)=w \\ 
\exists t_{f}\in \mathbf{R},\forall u\in U,\forall x\in f(u)\cap
S_{c}^{(n)},\forall t\geq t_{f},x(t)=x(t_{f})%
\end{array}%
\right. \Longleftrightarrow
\end{equation*}%
\begin{equation*}
\Longleftrightarrow \exists t_{f}\in \mathbf{R},\forall u\in U,\exists w\in 
\mathbf{B}^{n},\forall x\in f(u),\forall t\geq t_{f},x(t)=w
\end{equation*}

g) absolute constant stability with unbounded final time:%
\begin{equation*}
\left\{ 
\begin{array}{c}
\exists w\in \mathbf{B}^{n},\forall u\in U,\forall x\in f(u),\exists
t_{f}\in \mathbf{R},\forall t\geq t_{f},x(t)=w \\ 
\forall u\in U,\forall x\in f(u)\cap S_{c}^{(n)},\exists t_{f}\in \mathbf{R}%
,\forall t\geq t_{f},x(t)=x(t_{f})%
\end{array}%
\right. \Longleftrightarrow
\end{equation*}%
\begin{equation*}
\Longleftrightarrow \exists w\in \mathbf{B}^{n},\forall u\in U,\forall x\in
f(u),\exists t_{f}\in \mathbf{R},\forall t\geq t_{f},x(t)=w
\end{equation*}

h) absolute constant stability with bounded final time:%
\begin{equation*}
\left\{ 
\begin{array}{c}
\exists w\in \mathbf{B}^{n},\forall u\in U,\forall x\in f(u),\exists
t_{f}\in \mathbf{R},\forall t\geq t_{f},x(t)=w \\ 
\forall u\in U,\exists t_{f}\in \mathbf{R},\forall x\in f(u)\cap
S_{c}^{(n)},\forall t\geq t_{f},x(t)=x(t_{f})%
\end{array}%
\right. \Longleftrightarrow
\end{equation*}%
\begin{equation*}
\Longleftrightarrow \exists w\in \mathbf{B}^{n},\forall u\in U,\exists
t_{f}\in \mathbf{R},\forall x\in f(u),\forall t\geq t_{f},x(t)=w
\end{equation*}

i) absolute constant stability with fix final time:%
\begin{equation*}
\left\{ 
\begin{array}{c}
\exists w\in \mathbf{B}^{n},\forall u\in U,\forall x\in f(u),\exists
t_{f}\in \mathbf{R},\forall t\geq t_{f},x(t)=w \\ 
\exists t_{f}\in \mathbf{R},\forall u\in U,\forall x\in f(u)\cap
S_{c}^{(n)},\forall t\geq t_{f},x(t)=x(t_{f})%
\end{array}%
\right. \Longleftrightarrow
\end{equation*}%
\begin{equation*}
\Longleftrightarrow \exists w\in \mathbf{B}^{n},\exists t_{f}\in \mathbf{R}%
,\forall u\in U,\forall x\in f(u),\forall t\geq t_{f},x(t)=w
\end{equation*}%
where $w$ and $t_{f}$ are constant (thus $\exists w,\exists t_{f}$ commute)

\textbf{Theorem} Let the system $f$ having the property that it is
non-anticipatory and with fix final time.

a) If $f$ is absolutely stable, then the set $\Sigma _{f}(u)$ depends on the
restriction $u_{|(-\infty ,t_{f}]}$ only.

b) In the case that $f$ is absolutely delay-insensitive, the limit $\underset%
{t\rightarrow \infty }{\lim }x(t)$ that is the same for all $x\in f(u)$
depends on $u_{|(-\infty ,t_{f}]}$ only.

c) If $f$ is absolutely constantly stable, $\underset{t\rightarrow \infty }{%
\lim }x(t)$ is the same for all $x\in f(u)$ and all $u\in U$.

\section{Relative stability}

\textbf{Definition} a) A system $f$ that satisfies%
\begin{equation*}
\forall u\in U\cap S_{c}^{(m)},\forall x\in f(u),\exists w\in \mathbf{B}%
^{n},\exists t_{f}\in \mathbf{R},\forall t\geq t_{f},x(t)=w
\end{equation*}%
where $w$ and $t_{f}$ depend on $x$ only (thus $\exists w,\exists t_{f}$
commute) is called relatively stable.

b) If the next property is true%
\begin{equation*}
\forall u\in U\cap S_{c}^{(m)},\exists w\in \mathbf{B}^{n},\forall x\in
f(u),\exists t_{f}\in \mathbf{R},\forall t\geq t_{f},x(t)=w
\end{equation*}%
then $f$ is called relatively race-free stable, or relatively
delay-insensitive.

c) $f$ is relatively constantly stable if%
\begin{equation*}
\exists w\in \mathbf{B}^{n},\forall u\in U\cap S_{c}^{(m)},\forall x\in
f(u),\exists t_{f}\in \mathbf{R},\forall t\geq t_{f},x(t)=w
\end{equation*}%
If $U\cap S_{c}^{(m)}=\emptyset $ we say that the previous stability
properties are trivially fulfilled and if $U\cap S_{c}^{(m)}\neq \emptyset $
that they are non-trivially fulfilled.

\textbf{Remark} Relative stability is analized similarly with the absolute
stability.

\section{Stability relative to a function}

\textbf{Definition} Let the Boolean function $F:\mathbf{B}^{m}\rightarrow 
\mathbf{B}^{n}$.

a) A system $f$ satisfying%
\begin{equation*}
\forall u\in U\cap S_{F,c}^{(m)},\forall x\in f(u),\exists w\in \mathbf{B}%
^{n},\exists t_{f}\in \mathbf{R},\forall t\geq t_{f},x(t)=w
\end{equation*}%
where $w$ and $t_{f}$ depend on $x$ only (thus $\exists w,\exists t_{f}$
commute) is called $F-$relatively stable (or stable relative to the function 
$F$).

b) If the next property holds%
\begin{equation*}
\forall u\in U\cap S_{F,c}^{(m)},\forall x\in f(u),\exists t_{f}\in \mathbf{R%
},\forall t\geq t_{f},x(t)=F(u(t_{f}))
\end{equation*}%
then $f$ is called $F-$relatively race-free stable, or $F-$ relatively
delay-insensitive (race-free stable relative to the funcion $F$,
delay-insensitive relative to the function $F$).

c) $f$ is $F-$relatively constantly stable if it is $F-$relatively race-free
stable and the function $F$ is constant:%
\begin{equation*}
\exists w\in \mathbf{B}^{n},\forall u\in U,\forall x\in f(u),\exists
t_{f}\in \mathbf{R},\forall t\geq t_{f},x(t)=w
\end{equation*}%
If $U\cap S_{F,c}^{(m)}=\emptyset $ the previous stability properties are
trivial and if $U\cap S_{F,c}^{(m)}\neq \emptyset $ they are non-trivial.

\textbf{Remarks} The stability of a system relative to a Boolean function is
similar with the other notions of stability. We observe that the notions of $%
F-$relative constant stability and respectively of absolutely constant
stability coincide, being at the same time a special case of $F-$relative
race-free stability.

We give in Figure 1 the existing connection between the nine types of
stability that were previously defined.\FRAME{ftbpF}{4.5524in}{1.2635in}{0pt%
}{}{}{leg.jpg}{\special{language "Scientific Word";type
"GRAPHIC";maintain-aspect-ratio TRUE;display "USEDEF";valid_file "F";width
4.5524in;height 1.2635in;depth 0pt;original-width 4.4996in;original-height
1.2289in;cropleft "0";croptop "1";cropright "1";cropbottom "0";filename
'leg.jpg';file-properties "XNPEU";}}

\section{Synchronous-like, monotonous and hazard-free transitions. The
fundamental mode}

\textbf{Definition} For $x\in S^{(n)}$ and the time instances $t^{\prime
}<t" $, the couple $(x(t^{\prime }),x(t"))$ is called transition; we say
that $x$ has a transition in the interval $[t^{\prime },t"]$ from the value $%
x(t^{\prime })$ to the value $x(t")$.

\textbf{Definition} By the transition $(x(t^{\prime }-0),x(t"-0))$ it is
understood any of the transitions $(x(t^{\prime }-\varepsilon
),x(t"-\varepsilon ))$, where $\varepsilon >0$ is taken sufficiently small
so that%
\begin{equation*}
\forall \xi \in (0,\varepsilon ],x(t^{\prime }-\xi )=x(t^{\prime }-0)
\end{equation*}%
\begin{equation*}
\forall \xi \in (0,\varepsilon ],x(t"-\xi )=x(t"-0)
\end{equation*}%
The interval on which this transition takes place is by definition any of
the intervals $[t^{\prime }-\xi ,t"-\xi ]$ with $\xi \in (0,\varepsilon ]$.

\textbf{Notations} The usual notations for the transitions $(x(t^{\prime
}),x(t"))$ and $(x(t^{\prime }-0),x(t"-0))$ are $x(t^{\prime })\rightarrow
x(t")$ and respectively $x(t^{\prime }-0)\rightarrow x(t"-0)$. The interval
on which $x(t^{\prime }-0)\rightarrow x(t"-0)$ takes place is noted $%
[t^{\prime }-0,t"-0]$.

\textbf{Definition} The next data is given: the system $f$, the input $u\in
U $, the state $x\in f(u)$ and the instants $t^{\prime }<t"$. In this case
the transition $x(t^{\prime })\rightarrow x(t")$ is also called transfer of $%
x$ under the input $u$ in the interval $[t^{\prime },t"]$ from the value $%
x(t^{\prime })$ to the value $x(t")$ and we say that $f$ transfers $x$ under
the input $u$ (it $u-$transfers $x$) in the interval $[t^{\prime },t"]$ from 
$x(t^{\prime })$ to $x(t")$.

Similarly for the transition $x(t^{\prime }-0)\rightarrow x(t"-0)$.

\textbf{Definition} a) We suppose that $w,w^{\prime }\in \mathbf{B}%
^{n},t_{0},t_{f}\in \mathbf{R}$ and $u\in U$ exist so that

a.i) $\forall x\in f(u),\forall t<t_{0},x(t)=w$

a.ii) $\forall x\in f(u),\forall t\geq t_{f},x(t-0)=w^{\prime }$

a.iii) $t_{0}<t_{f}$

Then $x(t_{0}-0)\rightarrow x(t_{f}-0)$ is a synchronous-like transition (or
transfer); we say that $f$ transfers synchronous-likely (any) $x$ under the
input $u$ in the interval $[t_{0}-0,t_{f}-0]$ from the value $w$ to the
value $w^{\prime }$.

b) We suppose that $w,w^{\prime }\in \mathbf{B}^{n},t_{f},t_{f}^{\prime }\in 
\mathbf{R}$ and $u,v\in U$ exist so that

b.i) $\forall x\in f(u),\forall t\geq t_{f},x(t-0)=w$

b.ii) $\forall y\in f(v),\forall t\geq t_{f}^{\prime },y(t-0)=w^{\prime }$

b.iii) $t_{f}<t_{f}^{\prime }$

b.iv) $u_{|(-\infty ,t_{f})}=v_{|(-\infty ,t_{f})}$

b.v) $\{x_{|(-\infty ,t_{f})}|x\in f(u)\}=\{y_{|(-\infty ,t_{f})}|y\in
f(v)\} $

If they are true, then $y(t_{f}-0)\rightarrow y(t_{f}^{\prime }-0)$ is a
synchronous-like transition (or transfer). We also say that the system $f$
transfers synchronous-likely (any) $y$ under the input $v=u\cdot \varphi
_{(-\infty ,t_{f})}\oplus v\cdot \varphi _{\lbrack t_{f},\infty )}$ in the
interval $[t_{f}-0,t_{f}^{\prime }-0]$ from the value $w$ to the value $%
w^{\prime }$.

c) All the synchronous-like transitions are these from a) and b).

\textbf{Remarks} The attribute 'synchronous-like' given to a transition $%
y(t_{f}-0)\rightarrow y(t_{f}^{\prime }-0)$ implies the fact that $%
y(t_{f}-0)=x(t_{f}-0)$ is a steady value of $x\in f(u)$ and $y(t_{f}^{\prime
}-0)$ is a steady value $y\in f(v)$. The initial value is the same for all $%
x\in f(u)$ and all $y\in f(v)$ and it is treated as a steady value. Things
happen as if the unique state $y$ switches with all the coordinates
simultaneously (synchronously), in discrete time, in the manner $%
y(k)=w,y(k+1)=w^{\prime },...$ On the other hand, the 'composition' of the
synchronous-like transitions is a synchronous-like transition: if $%
t_{f}<t_{f}^{\prime }<t_{f}^{"}$ and if $y(t_{f}-0)\rightarrow
y(t_{f}^{\prime }-0),$ $y(t_{f}^{\prime }-0)\rightarrow y(t_{f}^{"}-0)$ are
synchronous-like transitions, then $y(t_{f}-0)\rightarrow y(t_{f}^{"}-0)$ is
synchronous-like too.

\textbf{Definition} Let the system $f$ and the input $u$ having the property
of existence of an unbounded sequence $t_{0}<t_{1}<t_{2}<...$ so that $%
x(t_{k}-0)\rightarrow x(t_{k+1}-0)$ be synchronous-like for all $k\in 
\mathbf{N}$ and all $x\in f(u)$. We say that $f$ is, under the input $u$, in
the fundamental (operating) mode.

\textbf{Definition} The non-empty set $U\subset S^{(m)}$ is called $\sigma -$%
closed if for any sequence $u^{k}\in U,k\in \mathbf{N}$ of inputs and any
unbounded sequence $t_{0}<t_{1}<t_{2}<...$ of real numbers we have $%
u^{0}\cdot \varphi _{(-\infty ,t_{0})}\oplus u^{1}\cdot \varphi _{\lbrack
t_{0},t_{1})}\oplus u^{2}\cdot \varphi _{\lbrack t_{1},t_{2})}\oplus ...\in
U $.

\textbf{Theorem} We suppose that $U$ is $\sigma -$closed and that $f$
satisfies

a) it is non-anticipatory

b) it satisfies the next property of initialization with bounded initial
time:%
\begin{equation*}
\forall u\in U,\exists w^{0}\in \mathbf{B}^{n},\exists t_{0}\in \mathbf{R}%
,\forall x\in f(u),\forall t<t_{0},x(t)=w^{0}
\end{equation*}%
where $w^{0}$ and $t_{0}$ depend on $u$ only (thus $\exists w^{0},\exists
t_{0}$ commute)

c) it is absolutely race-free stable with bounded final time, i.e.%
\begin{equation*}
\forall u\in U,\exists w\in \mathbf{B}^{n},\exists t_{f}\in \mathbf{R}%
,\forall x\in f(u),\forall t\geq t_{f},x(t-0)=w
\end{equation*}%
where $w$ and $t_{f}$ depend on $u$ only (thus $\exists w,\exists t_{f}$
commute)

Then for any sequence $u^{k}\in U,k\in \mathbf{N}$ of inputs, the unbounded
sequence $t_{0}<t_{1}<t_{2}<...$ of real numbers exists so that the
transitions $x(t_{k}-0)\rightarrow x(t_{k+1}-0),$ $k\in \mathbf{N},$ $x\in
f(u)$ are synchronous-like, where $u\in U$ is given by%
\begin{equation*}
u=u^{0}\cdot \varphi _{(-\infty ,t_{1})}\oplus u^{2}\cdot \varphi _{\lbrack
t_{1},t_{2})}\oplus u^{3}\cdot \varphi _{\lbrack t_{2},t_{3})}\oplus ...
\end{equation*}

\textbf{Remark} The previous theorem has two variants when '$f$ is
absolutely race-free stable' is replaced by '$f$ is relatively race-free
stable', respectively by '$f$ is $F-$relatively race-free stable'.

\textbf{Theorem} Let the system $f:U\rightarrow P^{\ast }(S^{(n)}),$ with $U$
$\sigma -$closed and we make the next suppositions:

a) $f$ is non-anticipatory

b) it is initialized with fix initial time, i.e.%
\begin{equation*}
\exists w^{0}\in \mathbf{B}^{n},\exists t_{0}\in \mathbf{R},\forall u\in
U,\forall x\in f(u),\forall t<t_{0},x(t)=w^{0}
\end{equation*}%
where $w^{0}$ and $t_{0}$ are constant (thus $\exists w^{0},\exists t_{0}$
commute)

c) \ the next controllability properties hold:%
\begin{equation}
\forall w\in \mathbf{B}^{n},\exists t_{f}\in \mathbf{R},\exists u\in
U,\forall x\in f(u),\forall t\geq t_{f},x(t-0)=w
\end{equation}%
\begin{equation}
\exists w\in \mathbf{B}^{n},\exists t_{f}\in \mathbf{R},\exists u\in
U,\forall x\in f(u),\forall t\geq t_{f},x(t-0)=w\Longrightarrow
\end{equation}%
\begin{equation*}
\Longrightarrow \forall w^{\prime }\in \mathbf{B}^{n},\exists t_{f}^{\prime
}\in \mathbf{R},\exists v\in U,\forall x\in f(u\cdot \varphi _{(-\infty
,t_{f})}\oplus v\cdot \varphi _{\lbrack t_{f},\infty )}),
\end{equation*}%
\begin{equation*}
\forall t\geq t_{f}^{\prime },x(t-0)=w^{\prime }
\end{equation*}%
Then for any sequence $w^{k}\in \mathbf{B}^{n},k\geq 1$ of binary vectors,
an unbounded sequence $t_{0}<t_{1}<t_{2}<...$ of real numbers and a sequence
of inputs $u^{k}\in U,k\in \mathbf{N}$ exist so that the input $u\in U$
defined by%
\begin{equation*}
u=u^{0}\cdot \varphi _{(-\infty ,t_{1})}\oplus u^{2}\cdot \varphi _{\lbrack
t_{1},t_{2})}\oplus u^{3}\cdot \varphi _{\lbrack t_{2},t_{3})}\oplus ...
\end{equation*}%
satisfies the property%
\begin{equation*}
x(t_{k}-0)=w^{k}
\end{equation*}%
\begin{equation*}
x(t_{k}-0)\rightarrow x(t_{k+1}-0)\quad are\quad synchronous-like
\end{equation*}%
for all $k\in \mathbf{N}$ and all $x\in f(u).$

\textbf{Definition} The transition $x(t^{\prime })\rightarrow x(t")$ is
called monotonous, if all the coordinate functions $x_{i},i=\overline{1,n}$
restricted to the interval $[t^{\prime },t"]$ are monotonous, i.e. they have
on $[t^{\prime },t"]$ at most one discontinuity point. The transition $%
x(t^{\prime }-0)\rightarrow x(t"-0)$ is monotonous if all the coordinate
functions $x_{i},i=\overline{1,n}$ restricted to all the intervals $%
[t^{\prime }-\varepsilon ,t"-\varepsilon ]$ with $\varepsilon >0$ chosen
sufficiently small are monotonous.

\textbf{Definition} If for $u\in U$ and $t_{f}<t_{f}^{\prime }$ the transfer 
$x(t_{f}-0)\rightarrow x(t_{f}^{\prime }-0)$ is synchronous-like and
monotonous, $x\in f(u)$ then it is called hazard-free.

\section{Conclusions}

The asynchronous systems are a mathematical concept that is inspired by the
modeling of the asynchronous circuits and the purpose of this paper is that
of stating the stability problem for them. We can furthermore connect with
this topic the notions of controllability and accessibility (by analogy we
can adopt from [1] about eight definitions of controllability and four
definitions of accessibility, but there exist also different points of view
in the literature) we can change / replace the non-anticipation condition
with other similar or dual conditions, we can suppose that $f$ is generated
by a generator function $\Phi :\mathbf{B}^{n}\times \mathbf{B}%
^{m}\rightarrow \mathbf{B}^{n}$, that it satisfies supplementary inertial
properties etc.

\end{document}